\def\r15{r^{-1.5}}
\def\r10{r^{-1.0} }
\def\r20{r^{-2.0} }

\def\micron{$\mu$m}
\def\l1551{L1551\ IRS\ 5}


\def\deg{$^{\circ}$}

\def\far-ir{far-infrared}


\def\subsubsection#1{\global\advance\count12 by 1
     \vskip \parskip \vskip \dimen2
     \centerline{{\number\count12}) {#1}}
     \vskip \dimen2}
%
%
\def\refindent{\advance\leftskip by 24pt \parindent=-24pt}
%
\def\journal#1#2#3#4#5{{\refindent
                      {#1}        
                      {#2},       
                      {\it #3\/}, 
                      {\bf #4},   
                      {#5}.       
                      \par }}
%

%
\def\inbook#1#2#3#4#5#6#7{{\refindent
                         {#1}         
                         {#2},        
                      in {\it #3\/},  
                     ed. {#4}         
                        ({#5}:        
                         {#6}),       
                       p.{#7}.        
                         \par }}
%

%

%

%

%


\expandafter \def \csname EQLABELEnergy.outflow\endcsname {0.1?}
\expandafter \def \csname EQLABELradius.pn.disk\endcsname {0.2?}
\expandafter \def \csname EQLABELvrot.pn.disk\endcsname {0.3?}
\expandafter \def \csname EQLABELr2.als\endcsname {0.4?}
\expandafter \def \csname EQLABELRc.als\endcsname {0.5?}
\expandafter \def \csname EQLABELdensity.als\endcsname {0.6?}
\expandafter \def \csname EQLABELdelta\endcsname {0.7?}
\expandafter \def \csname EQLABELchi.sq\endcsname {0.8?}
\expandafter \def \csname EQLABELtau.lambda\endcsname {0.9?}
\expandafter \def \csname TABLABELmodel.parameters\endcsname {0.1?}
\expandafter \def \csname TABLABELgradient.models\endcsname {0.2?}
\expandafter \def \csname TABLABELaverage.density\endcsname {0.3?}
\expandafter \def \csname TABLABELsummary.observations\endcsname {0.4?}
\expandafter \def \csname TABLABELdust.models\endcsname {0.5?}
\expandafter \def \csname FIGLABELscan.direction\endcsname {0.1?}
\expandafter \def \csname FIGLABELxdir.85\endcsname {0.2?}
\expandafter \def \csname FIGLABELydir.85\endcsname {0.3?}
\expandafter \def \csname FIGLABELzdir.87\endcsname {0.4?}
\expandafter \def \csname FIGLABELals.scan\endcsname {0.5?}
\expandafter \def \csname FIGLABELg1.model\endcsname {0.6?}
\expandafter \def \csname FIGLABELresiduals.models\endcsname {0.7?}
\expandafter \def \csname FIGLABELwalker.scan\endcsname {0.8?}
\expandafter \def \csname FIGLABELals.g1.sed\endcsname {0.9?}
\expandafter \def \csname FIGLABELg3.sed\endcsname {0.10?}
\expandafter \def \csname FIGLABELbeam.size.effect\endcsname {0.11?}
\expandafter \def \csname FIGLABELdust.opacity\endcsname {0.12?}
\expandafter \def \csname FIGLABELdiff.dust.models\endcsname {0.13?}
%
%
%

\newif\ifproofmode		
\proofmodefalse			

\newif\ifforwardreference	
\forwardreferencetrue 		

\newif\ifchapternumbers		
\chapternumberstrue		

\newif\ifcontinuousnumbering	
\continuousnumberingfalse	

\newif\iffigurechapternumbers	
\figurechapternumberstrue 	

\newif\ifcontinuousfigurenumbering	
\continuousfigurenumberingfalse	        

\font\eqsixrm=cmr5 scaled\magstephalf		
\def\marginstyle{\eqsixrm}	

\newtoks\chapletter		
\newcount\chapno		
\newcount\eqlabelno		
\newcount\figureno		
\newcount\tableno		

\chapno=0
\eqlabelno=0
\figureno=0
\tableno=0

%
\def\chapfolio{\ifnum \chapno>0 \the\chapno \else \the\chapletter \fi}

%
\def\bumpchapno{\ifnum \chapno>-1 \global \advance \chapno by 1
	\else \global \advance \chapno by -1 \setletter\chapno \fi
	\ifcontinuousnumbering \else \global\eqlabelno=0 \fi
	\ifcontinuousfigurenumbering \else \global\figureno=0 \tableno=0 \fi}

%
\def\setletter#1{\ifcase-#1 {}\or{} \or\global\chapletter={A}
  \or\global\chapletter={B} \or\global\chapletter={C} 
  \or\global\chapletter={D} \or\global\chapletter={E} 
  \or\global\chapletter={F} \or\global\chapletter={G} 
  \or\global\chapletter={H} \or\global\chapletter={I} 
  \or\global\chapletter={J} \or\global\chapletter={K} 
  \or\global\chapletter={L} \or\global\chapletter={M} 
  \or\global\chapletter={N} \or\global\chapletter={O} 
  \or\global\chapletter={P} \or\global\chapletter={Q} 
  \or\global\chapletter={R} 
  \or\global\chapletter={S} \or\global\chapletter={T} 
  \or\global\chapletter={U} \or\global\chapletter={V} 
  \or\global\chapletter={W} \or\global\chapletter={X} 
  \or\global\chapletter={Y} \or\global\chapletter={Z} \fi}

%
\def\tempsetletter#1{\ifcase-#1 {}\or{} \or\chapletter={A}
  \or\chapletter={B} \or\chapletter={C} \or\chapletter={D} \or\chapletter={E}
  \or\chapletter={F} \or\chapletter={G} \or\chapletter={H} \or\chapletter={I}
  \or\chapletter={J} \or\chapletter={K} \or\chapletter={L} \or\chapletter={M}
  \or\chapletter={N} \or\chapletter={O} \or\chapletter={P} \or\chapletter={Q}
  \or\chapletter={R} \or\chapletter={S} \or\chapletter={T} 
  \or\chapletter={U} \or\chapletter={V} \or\chapletter={W} \or\chapletter={X}
  \or\chapletter={Y} \or\chapletter={Z} \fi}

%
\def\chapshow#1{\ifnum #1>0 \relax #1%
   \else {\tempsetletter{\number#1}\chapno=#1 \chapfolio}\fi}

%
\def\today{\number\day\space \ifcase\month\or Jan\or Feb\or
	Mar\or Apr\or May\or Jun\or Jul\or Aug\or Sep\or 
	Oct\or Nov\or Dec\fi \space \number\year}

%
\def\initialeqmacro{\ifproofmode
	\headline{\tenrm \today\hfill \jobname\ --- draft\hfill\folio}
	 	\hoffset=-1cm \immediate \openout2=allcrossreferfile \fi
	\ifforwardreference 
		\ifproofmode \immediate\openout1=labelfile \fi \fi}


%
%

\def\chaplabel#1{\bumpchapno \ifproofmode \ifforwardreference
   \immediate\write1{\noexpand\expandafter\noexpand\def
   \noexpand\csname CHAPLABEL#1\endcsname{\the\chapno}}\fi\fi
   \global\expandafter\edef\csname CHAPLABEL#1\endcsname
   {\the\chapno}\ifproofmode\llap{\hbox{\marginstyle #1\ }}\fi\chapfolio}

%
%
\def\eqnum{\global\advance\eqlabelno by 1
	\eqno(\ifchapternumbers\chapfolio.\fi\the\eqlabelno)}

\def\eqlabel#1{\global\advance\eqlabelno by 1\ifproofmode \ifforwardreference
   \immediate\write1{\noexpand\expandafter\noexpand\def
   \noexpand\csname EQLABEL#1\endcsname{\the\chapno.\the\eqlabelno?}}\fi\fi
   \global\expandafter\edef\csname EQLABEL#1\endcsname
   {\the\chapno.\the\eqlabelno?} \eqno(\ifchapternumbers\chapfolio.\fi
   \the\eqlabelno)\ifproofmode\rlap{\hbox{\marginstyle\ #1}}\fi}

\def\eqalignlabel#1{\global\advance\eqlabelno by 1\ifproofmode 
   \ifforwardreference\immediate\write1{\noexpand\expandafter\noexpand\def
   \noexpand\csname EQLABEL#1\endcsname{\the\chapno.\the\eqlabelno?}}\fi\fi
   \global\expandafter\edef\csname EQLABEL#1\endcsname
   {\the\chapno.\the\eqlabelno?}&(\ifchapternumbers\chapfolio.\fi
   \the\eqlabelno)\ifproofmode\rlap{\hbox{\marginstyle\ #1}}\fi}

\def\eqref#1{(\ifundefined{EQLABEL#1}***\ifproofmode\ifforwardreference)%
   \else \write16{ ***Undefined Equation Reference #1*** }\fi
   \else \write16{ ***Undefined Equation Reference #1*** }\fi
   \else \edef\LABxx{\getlabel{EQLABEL#1}}%
   \chapshow{\LAByy.\expandafter\stripchap\LABxx}%
   \ifchapternumber\chapshow{\LAByy}.\expandafter\stripeq\LABxx%
   \else\ifnum \number\LAByy=\chapno \relax\expandafter\stripeq\LABxx
   \else\chapshow{\LAByy}.\expandafter\stripseq\LABxx\fi)\fi%
   \ifproofmode\write2{Equation #1}\fi}

%
%
\def\fignum{\global\advance\figureno by 1 \relax
   \iffigurechapternumbers\chapfolio.\fi\the\figureno}

\def\figlabel#1{\global\advance\figureno by 1\relax
   \ifproofmode \ifforwardreference
   \immediate\write1{\noexpand\expandafter\noexpand\def
   \noexpand\csname FIGLABEL#1\endcsname{\the\chapno.\the\figureno?}}\fi\fi
   \global\expandafter\edef\csname FIGLABEL#1\endcsname%
   {\the\chapno.\the\figureno?}
   \ifproofmode\llap{\hbox{\marginstyle #1\ }}
   \relax\fi\noindent 
          {\bf Figure~\ifchapternumbers\chapfolio.\fi\the\figureno:  }}

\def\figref#1{\ifundefined{FIGLABEL#1}!!!\ifproofmode\ifforwardreference%
   \else \write16{ ***Undefined Figure Reference #1*** }\fi
   \else \write16{ ***Undefined Figure Reference #1*** }\fi
   \else \edef\LABxx{\getlabel{FIGLABEL#1}}%
   \def\LAByy{\expandafter\stripchap\LABxx}%
   \iffigurechapternumbers\chapshow{\LAByy}.\expandafter\stripeq\LABxx
   \else\chapshow{\LAByy} \expandafter\stripeq\LABxx\fi\fi
   \ifproofmode\write2{Fig: #1}\fi}

%
%
\def\tabnum{\global\advance\tableno by 1 \relax
   \iffigurechapternumbers\chapfolio.\fi\the\tableno}

\def\tablabel#1{\global\advance\tableno by 1\relax
   \ifproofmode \ifforwardreference
   \immediate\write1{\noexpand\expandafter\noexpand\def
   \noexpand\csname TABLABEL#1\endcsname{\the\chapno.\the\tableno?}}\fi\fi
   \global\expandafter\edef\csname TABLABEL#1\endcsname
   {\the\chapno.\the\tableno?} 
   \ifproofmode\llap{\hbox{\marginstyle #1\ }}
   \relax\fi\centerline{\bf Table~\ifchapternumbers\chapfolio.\fi\the\tableno}%
    }

\def\tabref#1{\ifundefined{TABLABEL#1}!!!\ifproofmode\ifforwardreference%
   \else \write16{ ***Undefined table Reference #1*** }\fi
   \else \write16{ ***Undefined table Reference #1*** }\fi
   \else \edef\LABxx{\getlabel{TABLABEL#1}}%
   \def\LAByy{\expandafter\stripchap\LABxx}%
   \iffigurechapternumbers\chapshow{\LAByy}.\expandafter\stripeq\LABxx
   \else\chapshow{\LAByy} \expandafter\stripeq\LABxx\fi\fi%
   \ifproofmode\write2{Tab: #1}\fi}

%
%


%
%
\def\getlabel#1{\csname#1\endcsname}
\def\ifundefined#1{\expandafter\ifx\csname#1\endcsname\relax}
\def\stripchap#1.#2?{#1}
\def\stripeq#1.#2?{#2}

\font\twelverm = cmr12 \font\tenrm = cmr10
       \font\sevenrm = cmr7
\font\twelvei = cmmi12
       \font\teni = cmmi10 \font\seveni = cmmi7
\font\twelveit = cmti12 
       
\font\twelvesy = cmsy10 scaled\magstep1 
       \font\tensy = cmsy10 \font\sevensy = cmsy7
\font\twelvebf = cmbx12 \font\tenbf = cmbx10
       \font\sevenbf = cmbx7
\font\twelvesl = cmsl12
\font\twelveit = cmti12
\font\twelvett = cmtt12
%
\textfont0 = \twelverm \scriptfont0 = \tenrm 
       \scriptscriptfont0 = \tenrm
       \def\rm{\fam0 \twelverm}
\textfont1 = \twelvei \scriptfont1 = \teni 
       \scriptscriptfont1 = \teni
       
\textfont2 = \twelvesy \scriptfont2 = \tensy 
       \scriptscriptfont2 = \tensy
       
\newfam\itfam \def\it{\fam\itfam \twelveit} \textfont\itfam=\twelveit
\newfam\slfam  \textfont\slfam=\twelvesl
\newfam\bffam \def\bf{\fam\bffam \twelvebf} \textfont\bffam=\twelvebf
       \scriptfont\bffam=\twelvebf \scriptscriptfont\bffam=\tenbf
\newfam\ttfam  \textfont\ttfam=\twelvett
\rm
\hsize=6in
\hoffset=.45in
\vsize=9in
\baselineskip=24pt
%
\raggedright  \pretolerance = 800  \tolerance = 1100
\raggedbottom
%
\dimen1=\baselineskip \multiply\dimen1 by 3 \divide\dimen1 by 4
\dimen2=\dimen1 \divide\dimen2 by 2
%
\def\apjsingle{\baselineskip = 14pt 
               \parskip=14pt plus 1pt
               \dimen1=\baselineskip \multiply\dimen1 by 3 \divide\dimen1 by 4
               \dimen2=\dimen1 \divide\dimen2 by 2
               \scriptfont0 = \tenrm  \scriptscriptfont0 = \sevenrm
               \scriptfont1 = \teni  \scriptscriptfont1 = \seveni
               \scriptfont2 = \tensy \scriptscriptfont2 = \sevensy
               \scriptfont\bffam = \tenbf \scriptscriptfont\bffam = \sevenbf
               \rightskip=0pt  \spaceskip=0pt  \xspaceskip=0pt
               \pretolerance=100  \tolerance=200
              }
%
\nopagenumbers
\headline={\ifnum\pageno=1 \hss\thinspace\hss 
     \else\hss\folio\hss \fi}
%
 
%
\count10 = 0
\def\section#1{\vbox to \dimen1 {\vfill}
    \global\advance\count10 by 1
    \centerline{\uppercase\expandafter{\number\count10}.\ \bf {#1}}
    \global\count11 = 0  
    \vskip \dimen1}
%
\def\subsection#1{\global\advance\count11 by 1 
    \vskip \parskip  \vskip \dimen2
    \centerline{\uppercase{\number\count10}.\uppercase{\number\count11}.
    \ \it {#1}} 
    \global\count12=0 
    \vskip \dimen2}
\def\subsubsection#1{\global\advance\count12 by 1
     \vskip \parskip \vskip \dimen2
     \centerline{{\number\count10}.{\number\count11}.{\number\count12}
     \ {#1}}
     \vskip \dimen2}

%
%
\def\refindent{\advance\leftskip by 24pt \parindent=-24pt}
%
\def\journal#1#2#3#4#5{{\refindent
                      {#1}        
                      {#2},       
                      {\it #3},       
                      {\bf #4},       
                      {#5}.       
                      \par }}
%

%

%
\def\inbook#1#2#3#4#5#6#7{{\refindent
                         {#1}         
                         {#2},        
                      in {\it #3\/},  
                     ed. {#4}         
                        ({#5}:        
                         {#6}),       
                       p.{#7}.        
                         \par }}
%

%


%

%

%

%

%

%
\def\etal{{\it et al.\/\ }}

\proofmodefalse
\forwardreferencetrue
\initialeqmacro
\chapternumbersfalse
\figurechapternumbersfalse

\centerline{  }
\bigskip\bigskip
\bigskip\bigskip
\bigskip\bigskip
\bigskip\bigskip
\bigskip\bigskip
\hoffset=.2in
\centerline{\bf Far-Infrared Constraints on Dust Shells}
\centerline{\bf Around Vega-Like Stars}
\bigskip
\bigskip
\centerline{Paul M. Harvey, Beverly J. Smith, James DiFrancesco, Cecilia Colom\'e}
\centerline{Department of Astronomy, The University of Texas at Austin}
\bigskip\bigskip
\centerline{and Frank J. Low, Steward Observatory, University of Arizona}
\bigskip\bigskip
\bigskip\bigskip
\bigskip\bigskip
\bigskip\bigskip
\bigskip\bigskip
\bigskip\bigskip
\bigskip\bigskip
\hskip 1.4in{{\bf Received} \vbox{\hskip 3cm \hrule width4cm}}
\bigskip \bigskip \bigskip \bigskip \bigskip \bigskip \bigskip
\bigskip\bigskip
\hskip 1.4in{\bf Astrophysical Journal, in press}
\vfill \eject

\centerline{\bf Abstract}

We present results of observations at 47 and 95 $\mu$m from the Kuiper
Airborne Observatory of several ``Vega-like'' stars.  Spatial cuts and
aperture photometry are presented for $\beta$ Pictoris, Fomalhaut, and
HD 135344, 139614, 142527, and 169142, four stars that had been suggested
to possibly represent more distant examples of the Vega phenomenon by
Walker and Wolstencroft.  We have modelled the dust around
$\beta$ Pic and Fomalhaut with a spatially and optically thin disk to
determine the constraints our new observations place on the properites
of the dust disks that are required to explain the infrared and optical
properties of these two stars.  For $\beta$ Pic we find that models similar
to those proposed by Backman, Gillett, and Witteborn can fit our data
quite well.  For Fomalhaut we find that very different models are required
which have much ``blacker'' dust with a much shallower density distribution,
surface density $\propto r^{-0.5}$, than for $\beta$ Pic.
Our observations of the four HD stars are consistent with their being
spatially unresolved.  Because of their distance, this does not allow us
to put any new constraints on their circumstellar shells.

\vfill \eject

\section{Introduction}

One of the most exciting and unexpected discoveries made by the IRAS survey
was that of thermal dust emission from main-sequence stars, the ``Vega''
phenomenon (Aumann \etal 1984; Gillett 1986).  Not only was dust unexpected
around main sequence stars that were not undergoing detectable mass loss,
but the cloud sizes and temperatures indicated
that the particles must be much larger than typical interstellar
grains.  Furthermore, estimates of the lifetimes of such grains against
radiation pressure and the Poynting-Robertson effect showed that the 10 to
100 \micron\ diameter grains probably had to be constantly replenished from a
reservoir of, perhaps, much larger grains.  
Unfortunately the spatial resolution of IRAS in the far-infrared,
where these clouds are most luminous, was barely sufficient to resolve them.
Therefore, the constraints on the dust properties due to the spatial extent
of the clouds were not completely defined.  

Immediately after the announcement of this discovery, the Kuiper Airborne
Observatory was used to make both confirming observations of the phenomenon
and to supply additional spatial and spectral constraints on the emission
(Harvey, Wilking, and Joy 1984 (HWJ); Harper, Lowenstein, and Davidson 1984).
These results as
well as a recent re-analysis of the HWJ data (van der Bliek, Prusti, and
Waters 1994) showed the
value of the substantially higher spatial resolution possible with the KAO,
even though its sensitivity is much poorer than that of IRAS.  

For the past half-dozen years we have been attempting to obtain the highest
possible quality new observations of several of these objects with the KAO.
Since the largest fraction of them lie at southern declinations, this has
involved the use of the KAO on its regular deployments to New Zealand
which began with the appearance of SN1987a.  In this paper we report the
results of these observations and discuss their implications for the
properties of the circumstellar clouds around the observed stars.  Our
most complete data are on $\beta$ Pictoris and Fomalhaut ($\alpha$ Piscis Austrinus).
In addition, we have obtained some observations on four stars that were
suggested by Walker and Wolstencroft (1988) to have properties similar to
the four original stars found by IRAS.  In the following sections of this
paper we discuss: the details of our observations, the basic observational
results, simple models that can fit most of the observational data 
on $\beta$ Pic and Fomalhaut, and constraints on the dust shells around
the additional stars.

\section {Observations}

All the observations presented here were made on the KAO flying out of
Christchurch, New Zealand,  between 1988 and
1994 with one of two detector systems.  The first system consisted of
an array of 1 $\times$ 8 bolometers imaged on the focal plane with a pixel
scale of roughly $\lambda /2D \times \lambda/D$ at 
effective wavelengths of either 47 or 95 \micron\
($\lambda/\Delta\lambda \sim 1.5$)(Smith \etal 1991).
The second system used the same optics, filtering, and pixel scale with 
a 2 $\times$ 10
array of bolometers (Smith \etal 1994).  The details of the observations, 
including dates, calibration
objects, detector system used, wavelengths, and objects observed are
given in Table 1.  Also shown in Table 1 is the rotational orientation of
the sky relative to the detector arrays; for these angles the convention
is that the long axis of each array was along the 0\deg - 180\deg\ line
and the angle indicated is that by which the sky was apparently rotated
from north (0\deg) through west (90\deg).  All of the observed objects are bright, visible
stars, so no off-axis guiding was required.  The absolute calibrations
are believed accurate to $\pm$15\%, except in the case of the 1992 data
where the large number and consistency of calibration sources gives
calibration uncertainties of $\pm$10\%.

\section {Observational Results}

The results for all the objects observed are shown in Figures
1~-~4 and Table 2.  The figures illustrate the spatial results and relevance of
our flux density measurements to the overall energy distributions, and the table lists
peak flux densities in our KAO beams relative to the IRAS large-beam results.
Several of the figures also include modeling results that will be discussed
below.

For $\beta$ Pic it was impossible to schedule flights at a time when the orientation
of our array would be along the major axis of the circumstellar disk.  Therefore,
our observations consist of peak flux measurements with the central pixel of the
array (Table 2) together with measurements  of
several points 1/2 and 1 beamwidth on either side of the star along a line
through the optical circumstellar disk (Figure 1, where
all the measurements on either side have been averaged to improve the 
signal-to-noise ratio).  The peak flux data (Table 2) show that 
at 95 \micron\ the circumstellar cloud is marginally resolved
in our KAO beam relative to the IRAS beam; at 47 \micron\ there is
clear evidence with our much smaller KAO beam that the cloud is resolved
in the peak flux data; the 47 \micron\ spatial data are also consistent with
the idea that
the source has been resolved, though they are certainly not compelling.

Our Fomalhaut data are summarized in Figure 3 and Table 2.  The 1988 data are the
only observations made with the long axis of our array essentially parallel to the
IRAS in-scan direction along which the source appeared to be resolved by IRAS.  All of the
other 95 \micron\ observations, as well as the only 47 \micron\ observations,
were made with the array roughly parallel to the IRAS cross-scan direction, along
which Fomalhaut showed no evidence for resolution by IRAS.  As noted in a preliminary report
on our first observations by Lester \etal (1988), our spatial data along the IRAS
in-scan direction appear to resolve the circumstellar cloud at 95 \micron.  In the
perpendicular direction (all our other data), there is some evidence for resolution
at 47 \micron, and slight evidence at 95 \micron.  In the comparison of the peak
flux measured in the KAO beam relative to the IRAS beams, however, Table 2 shows
strong evidence that Fomalhaut is resolved in all the KAO observations assuming
there has not been significant far-infrared variability over this time scale.

For the Walker and Wolstencroft objects all of the spatial data (Figure 4) are completely consistent
with their circumstellar clouds being pointlike at the KAO resolution.  
The flux density results
in Table 2, however, show that in two of the four observed cases, 
the 47 \micron\ KAO flux
measurements are lower than the IRAS fluxes by slightly more than the combined 1 $\sigma$ 
uncertainties in both.  We do not consider this strong evidence for resolution
of the clouds, but it suggests that there may be a small amount of emitting 
dust beyond the limits of the KAO beam.

\section {Modeling}

In order to compare our size information on these stars quantitatively with
the IRAS data and to determine the implications of our data for the
properties of the circumstellar clouds, we have tried fitting simple
models to the available data.  These models were designed to be similar
to those that Backman, Gillett, and Witteborn (1992; hereafter BGW) 
constructed for $\beta$ Pic,
and we have extended them to Fomalhaut as well.  The basic structure
of the models assumes a spatially and optically thin disk.  For $\beta$ Pic
the disk was assumed to be edge-on as suggested by the optical imagery; for
Fomalhaut the inclination angle was allowed to vary.  The circumstellar dust
was assumed to be distributed in a single power-law distribution between an
inner and outer radius, or, for the most successful $\beta$ Pic models,
to have an inner distribution with one power-law, optical depth, and emissivity
law, together with an outer distribution with different combination of
power-law/$\tau$/emissivity.   In these latter, two-component models, the
radius dividing the two regions was fixed at the value used by BGW of 80 AU.
The emissivity law for the dust in either
component was characterized in a simple way; shortward of a specified 
wavelength, $\lambda_\circ$,  the emissivity was assumed to be constant; longward of that
wavelength it was assumed to decrease as either $\epsilon \propto \lambda^{-1}$ or
$\epsilon \propto \lambda^{-2}$, although no models with a $\lambda^{-2}$
dependence provided good fits to any of the data.

\subsection{$\beta$ Pic}

Our aim in modeling $\beta$ Pic was, first, to reproduce the results of
BGW, and then to determine what additional constraints our data placed on
their model results.  BGW found that they were only able to fit the spectral
and spatial data on $\beta$ Pic with two-component models which had a
substantially lower dust density distribution inside a radius of order
100 AU.  They confirmed Gillett's conclusion that: (1) on average the grains
around $\beta$ Pic were smaller than those around Vega and Fomalhaut
(i.e., $\lambda_\circ \sim$ few \micron), and (2) the inner radius for any thermally emitting dust,
though model dependent, was of order 5 - 20 AU.

The additional constraints which we used from our data were the total flux
densities measured in our KAO beams and the spatial data at 47 \micron.
Also, subsequent to BGW's paper, Zuckerman and Becklin (1993)
published results of sub-mm photometry and mapping of
$\beta$ Pic and Fomalhaut.   We have also
included their 800 \micron\ data, although
small differences in the assumed emissivity law in the far-infrared can
make almost any of the realistic models fit the 800 \micron\ photometry.

The two best-fitting models of BGW were those labelled ``10'' and ``11'' in
their paper.  In both, the radius where the density changed from low
to high was 80 AU, and the surface density power-law was set at -1.7 in
the outer region to match that inferred from the optical coronagraphy.
The major differences in the two models were that model 10 used smaller
grains and a larger inner radius to reproduce the shorter wavelength
emission, and model 11 allowed a different power-law for the density
gradient in the inner disk region but assumed identical grain properties
in both regions.  Because these provided the best fits to the data available
to BGW, we attempted to fit our data as well as the previous data with
one or both of these two models.  Our results are shown in Figures 1 and 2
for a model we have labelled B10A which is quite similar to model 10 of BGW;
we also found a model similar to their model 11 (which we called B11A) that
gives comparable results which
are not shown here.  Figure 1 shows our spatial data relative to the model
prediction; Figure 2 shows the model energy
distribution, both total flux density, and flux density in various
aperture sizes, including the IRTF 4'' and 8'' mid-infrared data discussed by BGW,
our 47 and 95 \micron\ KAO data, and Zuckerman and Becklin's 800 \micron\
observations.  Both models 10A and 11A provide a reasonable fit
to all the data.  The parameters for these models are listed in
Table 3.
We also confirmed
that single density gradient models could not reproduce the observations.
Two different regions with densities differing by a factor of
15 - 20 are required to produce enough far-infrared emission without
overproducing the near-infrared emission and to fit the observed spatial
extent in the 10 - 20 \micron\ spectral region.

\subsection {Fomalhaut}

Gillett (1986) suggested simple models for the circumstellar clouds around
the four original Vega-like stars to explain the IRAS photometry and scan
data.  For Fomalhaut he found that a distribution of black grains with
a mild density gradient over a range of radii, 28 - 140 AU, gave a
reasonable fit to the IRAS data.  Therefore, we began our attempts to
fit the IRAS, KAO, and sub-mm data (Zuckerman and Becklin 1993) with
a distribution of grains with $\lambda_\circ \sim$ 100 \micron\ and a similar
density gradient and range of radii.  The facts that: (1) the IRAS in-scan
and cross-scan source sizes were clearly different, and (2) that
somewhat ``accidentally'' our KAO observations provided one dimensional
source profiles roughly along the same directions as the IRAS data, suggested
that we model the circumstellar cloud as an inclined disk.  With no additional
data or constraints on the disk inclination, nor the orientation of the
IRAS or KAO scans relative to the disk, we have fit the data assuming that
the IRAS in-scan data and 1988 KAO data were taken along the long axis of
the source and that the IRAS cross-scan and subsequent KAO data were
taken along the short apparent dimension.  Because of this free parameter and
the smaller quantity of spatial observations available than for $\beta$ Pic,
we concentrated our efforts on simple models with one dust density gradient
and one type of dust between an inner and outer radius.  This implies
six free parameters for these models: surface density power-law, inner and outer
radii, $\lambda_\circ$, optical depth, and inclination angle relative to the
line of sight.  For all the models we assumed a $\lambda^{-1}$ emissivity
law for $\lambda > \lambda_\circ$.

Figure 3 shows the results of the fits for the best model we found; the details
of this model as well as
models with other density gradients producing acceptable fits are listed in
Table 4.
The most important general features of models producing
acceptable fits are: (1) a large range of radii over which a substantial
amount of dust exists, (2) grains which have constant emissivity out
to $\lambda \sim$100 \micron, and (3) inclination angles between
45\deg\ and 75\deg.  (Because of the unknown orientation of the observations
relative to the supposed disk, these values represent lower limits to the
disk orientation).  Surface density gradients, $\sigma \propto r^{-n}$, with
$n \sim 0.5 \pm 0.5$ provide the only reasonably acceptable fits.
Steeper gradients put too much flux into the KAO beam relative to the
IRAS beam for models which reproduce the IRAS total fluxes; shallower
density gradients have the opposite problem (as well as being difficult
to understand on physical grounds).  A relatively large range in radii for
the dust is needed to explain the decrease in flux observed with the
KAO relative to IRAS; the derived range of inclination angles for the
model disks is required to fit the differences in flux density observed
in different array orientations on the KAO (as well as IRAS) and the
spatial KAO data.

\subsection{Walker--Wolstencroft Stars}

We have not performed any detailed modeling to fit the data on these objects for
the following reason.  For all these stars which we observed, simple-minded
models of circumstellar dust shells suggest that the shells should not be
resolvable at the KAO limit even for dust grains with properties comparable
to typical interstellar grains.  For example, if we assume a single temperature
dust shell whose temperature is determined by radiative equilibrium between
power absorbed with efficency, $\epsilon_a$, and power radiated with efficiency,
$\epsilon_r$, the calculated diameter of the circumstellar shells for these
four stars ranges from 3'' for HD169142 to 10'' for HD 142527 for a
ratio $\epsilon_a/\epsilon_r = 100$, typical of normal interstellar grains at
temperatures of $\sim$ 100K.  For blacker grains, the radius of a dust shell in
thermal equilibrium would only be smaller than the above sizes, so we have no
constraints on dust properties for grains in thermal equilibrium.
Clearly, this also implies that if IRAS
indeed resolved some of these dust shells, the extended emission must be due
to faint, low level, extended dust which is cooler than the bulk of the dust
contributing to the far-infrared fluxes reported by Walker and Wolstencroft (1988) 
(and confirmed by our KAO photometry).  Perhaps a small number of tiny grains
which are not in thermal equilibrium could explain the IRAS results.

\section{Conclusions}

The basic conclusion from our high-resolution work is that the models that
explain the lower resolution IRAS data for $\beta$ Pic and Fomalhaut are
quite consistent with our higher-resolution KAO observations.  Our most
important conclusion for $\beta$ Pic is that like BGW, we
find that its circumstellar disk can be well fitted with a two
component model whose main features are a substantially lower dust density
inside $\sim$ 100 AU and dust grains with a characteristic size of order
a few microns.  For Fomalhaut  our most important conclusions are, first
that we confirm Gillett's (1986) suggestion that the
grains must be essentially black out to the longest wavelengths observed by
IRAS and the KAO, 100 $\mu$m. Secondly, the dust density gradient around 
Fomalhaut is probably in the range of $\rho \propto r^{-0.5 \pm 0.5}$.  
In addition, we find some evidence that Fomalhaut's circumstellar
disk axis is likely to be inclined substantially to the line of sight, though
the data do not seem consistent with an angle as high as that of $\beta$ Pic.
These conclusions show that the disks around $\beta$ Pic and Fomalhaut are
different in a number of important ways.  In addition to Fomalhaut's being
substantially lower optical depth (as are all the other related stars), there
are large differences in particle size and in density gradient.

Our results on the four stars in Walker and Wolstencroft's list are difficult to
reconcile with their analysis of
the IRAS data which suggested that these stars have resolved dust
shells in the far-infrared at the IRAS resolution of $\sim$ 1 - 2'.  With
our KAO resolution of 10 - 20'', these stars should have been easily
resolved.  Instead, we found them to be essentially point-like, both in
the spatial cuts and by a simple comparison of KAO and IRAS flux densities.
The observed dust temperatures and assumption of
grain sizes even as small as typical interstellar grains do not
require the dust shells to be large enough to
be resolved by the KAO. 
Therefore, we cannot put any significant limits on the dust properties
around these stars.  On the other hand, the fact that we have not resolved them,
even though they appeared to be in the IRAS data, suggests that a careful
search for faint, extended emission by ISO would be worthwhile.

\centerline{Acknowledgments}

We thank the staff of the KAO for their superb support during the number 
of years
over which these observations were obtained. We hope that some of them will still
be on the airborne astronomy team when SOFIA begins flying to make the next
logical step in spatial resolution of these dust clouds.

\vfill
\eject
\centerline{\bf References}

\journal {Aumann, H., Gillett, F.C., Beichman, C., DeJong, T., Houck, J., Low, F., Neugebauer, G., Walker, R., and Wesselius, P.}{1984} {Ap.J. (Letters)}{278}{L23}
\journal {Backman, D.E., Gillett, F.C., and Witteborn, F.C.}{1992}{Ap.J.}{385}{670}
\inbook {Gillett, F.C.}{1986}{Light on Dark Matter}{Israel, F.}{Dordrecht}{Reidel}{61}
\journal {Harper, D.A., Loewenstein, R.F., and Davidson, J.A.}{1984}{Ap.J.}{285}{808}
\journal {Harvey, P.M., Wilking, B.A., and Joy, M.}{1984}{Nature}{307}{441}
\journal {Lester, D., Harvey, P., Smith, B., Colom\'e, C., and Low, F.}{1990}{B.A.A.S.}{21}{1085}
\journal {Smith, B.J., Lester, D.F., Harvey, P.M., and Pogge, R.W.}{1991}{Ap.J.}{373}{66}
\journal {Smith, B.J., Harvey, P.M., Colom\'e, C., Zhang, C.Y., Di Francesco, J., and Pogge, R.W.}{1994}{Ap.J.}{425}{91}
\journal {van der Bliek, N.S., Prusti, T., and Waters, L.B.F.M.}{1994}{Astr.Ap.}{285}{229}
\journal {Walker, H.J., and Wolstencroft, R.D.}{1988}{P.A.S.P.}{100}{1509}
\journal {Zuckerman, B. and Becklin, E.E.}{1993}{Ap.J.}{414}{793}

\vfill
\eject
	
\centerline{\bf Table 1}
\bigskip
\centerline{\bf Journal of Observations}
\bigskip
\bigskip
\halign{\hfil#\hfil&\quad\hfil#\hfil&\quad\hfil#\hfil&\quad\hfil#\hfil&\quad\hfil#\hfil&\quad\hfil#\hfil\cr
\noalign{\hrule}
\noalign{\smallskip}
\bf Date&\bf Calibrator&\bf Detectors& \bf $\lambda$ ($\mu$m)&\bf Star& \bf Field Rot. (\deg)\cr
\noalign{\smallskip}
\noalign{\hrule}
\noalign{\bigskip}
1988, Nov& Ceres, $\eta$ Car & 1 $\times$ 8& 95 & $\beta$ Pic & 290\cr
& & & 95 & Fomalhaut & 125\cr
\noalign{\bigskip}
1989, Apr & Uranus, $\eta$ Car& 1 $\times$ 8& 95& Fomalhaut& 240\cr
\noalign{\bigskip}
1990, May & Uranus, $\eta$ Car& 2 $\times$ 10& 95& Fomalhaut& 245\cr
\noalign{\bigskip}
1991, Apr& Ceres, $\eta$ Car& 2 $\times$ 10& 95 & Fomalhaut& 240\cr
\noalign{\bigskip}
1992, Mar& Uranus, $\eta$ Car,& 2 $\times$ 10& 47, 95& $\beta$ Pic& 115\cr
& Callisto, Ceres,\cr
& Neptune,\cr
& Ganymede\cr
\noalign{\bigskip}
1993, Apr/May& Uranus, $\eta$ Car& 2 $\times$ 10& 47, 95& Fomalhaut& 250\cr
& & & & HD135344& 95\cr
& & & & HD139614& 90\cr
\noalign{\bigskip}
1994, Jul/Aug& Uranus, $\eta$ Car& 2 $\times$ 10& 47, 95& Fomalhaut& 245\cr
& & & & HD135344& 105\cr
& & & & HD142527& 90\cr
& & & & HD169142& 125\cr
}
\vfill
\eject

\apjsingle{
\centerline{\bf Table 2}
\bigskip
\centerline{\bf Photometric Results}
\bigskip
\bigskip
\smallskip
\settabs\+&Fomalhautxxxx  & fx50uumxxxxx & XXXXXX& XXXXXXXXX& Fx100uumxxxxxxxxxxx \cr
\hrule
\smallskip
\+& {\bf STAR} & {\bf KAO F$_{\nu}$ (Jy)} & $\pm$ {\bf stat.} & @ $\lambda$ & {\bf IRAS F$_{\nu}$ (Jy)} @ $\lambda$\cr
\+&  &  & $\pm$ {\bf total} & \cr
\smallskip
\hrule
\smallskip
\+& $\beta$ Pic & \hfil 12.9 \hfil & $\pm$ 1.0 & @ 47$\mu$m & 18.8 $\pm$ 0.9 @ 60 $\mu$m\cr
\+& & & $\pm$ 1.6 & & \cr
\bigskip
\+&  & \hfil 8.5 \hfil & $\pm$ 0.6 & @ 95$\mu$m & 11.2 $\pm$ 1.0 @ 100 $\mu$m\cr
\+& & & $\pm$ 1.0 & & \cr
\bigskip
\bigskip
\+& Fomalhaut & \hfil 5.6 \hfil & $\pm$ 0.65 & @ 47$\mu$m & 9.8 $\pm$ 0.5 @ 60 $\mu$m\cr
\+& & & $\pm$ 0.95 & & \cr
\bigskip
\+&  & \hfil 6.7 \hfil & $\pm$ 0.6 & @ 95$\mu$m & 11.3 $\pm$ 1.1 @ 100 $\mu$m\cr
\+& & & $\pm$ 1.0 & & \cr
\bigskip
\bigskip
\+& HD 135344 & \hfil 24.3 \hfil & $\pm$ 1.8 & @ 47$\mu$m & 26.3 $\pm$ 1.5 @ 60 $\mu$m\cr
\+& & & $\pm$ 3.5 & & \cr
\bigskip
\bigskip
\+& HD 139614 & \hfil 14.0 \hfil & $\pm$ 1.3 & @ 47$\mu$m & 18.3 $\pm$ 1.2 @ 60 $\mu$m\cr
\+& & & $\pm$ 2.5 & & \cr
\bigskip
\bigskip
\+& HD 142527 & \hfil 98 \hfil & $\pm$ 2.0 & @ 47$\mu$m & 106 $\pm$ 6 @ 60 $\mu$m\cr
\+& & & $\pm$ 15 & & \cr
\bigskip
\+&  & \hfil 84 \hfil & $\pm$ 1.2 & @ 95$\mu$m & 82 $\pm$ 5 @ 100 $\mu$m\cr
\+& & & $\pm$ 12 & & \cr
\bigskip
\bigskip
\+& HD 169142 & \hfil 22.9 \hfil & $\pm$ 2.1 & @ 47$\mu$m & 28.9 $\pm$ 2 @ 60 $\mu$m\cr
\+& & & $\pm$ 3.5 & & \cr

\vfill
\eject

\centerline{\bf Table 3}
\bigskip
\centerline{\bf $\beta$ Pic Models}
\bigskip
\bigskip
$$\vbox{\halign{#\hfil&\quad\quad\hfil#\hfil&\quad\quad\hfil#\hfil\cr
\noalign{\hrule}
\noalign{\smallskip}
{\bf Parameter} & {\bf Model B10A} & {\bf Model B11A}\cr
\noalign{\smallskip}
\noalign{\hrule}
\noalign{\bigskip}
Incl. Angle & 90\deg & 90\deg\cr
\noalign{\smallskip}
Inner Radius & 20 a.u. & 5 a.u.\cr
\noalign{\smallskip}
Middle Radius & 80 a.u. & 80 a.u.\cr
\noalign{\smallskip}
Outer Radius & 2000 a.u. & 2000 a.u.\cr
\noalign{\smallskip}
$\gamma$ (inner)$^*$ & -1.7 & -0.4\cr
\noalign{\smallskip}
$\gamma$ (outer)$^*$ & -1.7 & -1.7\cr
\noalign{\smallskip}
$\lambda_o$ (inner)& 0.3 $\mu$m & 2.5 $\mu$m\cr
\noalign{\smallskip}
$\tau_{100}$ (inner)& 2.9 $\times$ 10$^{-4}$ & 2.8 $\times$ 10$^{-4}$\cr
\noalign{\smallskip}
$\tau_{100}$ (outer)& 5.1 $\times$ 10$^{-3}$ & 5.0 $\times$ 10$^{-3}$\cr
\noalign{\smallskip}
n$^{**}$ & -1 & -1\cr
}}$$
\vfil

\hskip 1.0in $^*$Surface density $\propto r^\gamma$;
$^{**}$Dust emissivity $\propto \lambda^n$
\vfill
\eject

\centerline{\bf Table 4}
\bigskip
\centerline{\bf Fomalhaut Models}
\bigskip
\bigskip
$$\vbox{\halign{#\hfil&\quad\quad\hfil#\hfil&\quad\quad\hfil#\hfil&\quad\quad\hfil#\hfil\cr
\noalign{\hrule}
\noalign{\smallskip}
{\bf Parameter} & {\bf Best Model, \#1} & {\bf Model \#2} & {\bf Model \#3}\cr
\noalign{\smallskip}
\noalign{\hrule}
\noalign{\bigskip}
Incl. Angle & 60\deg & 65\deg & 50\deg\cr
\noalign{\smallskip}
Inner Radius & 22 a.u. & 16 a.u. & 25 a.u.\cr
\noalign{\smallskip}
Outer Radius & 430 a.u. & 300 a.u. & 400 a.u.\cr
\noalign{\smallskip}
$\gamma^*$ & -0.5 & 0.0 & -0.75\cr
\noalign{\smallskip}
$\lambda_o$ & 80 $\mu$m & 80 $\mu$m & 85 $\mu$m\cr
\noalign{\smallskip}
$\tau_{100}$ & 2.7 $\times$ 10$^{-5}$ & 6.4 $\times$ 10$^{-5}$ & 5.6 $\times$ 10$^{-5}$\cr
\noalign{\smallskip}
n$^{**}$ & -1 & -1 & -1\cr
}}$$
\vfil

\hskip 1.0in $^*$Surface density $\propto r^\gamma$;
$^{**}$Dust emissivity $\propto \lambda^n$
\vfill
\eject
}
\baselineskip=24pt
\centerline{\bf Figure Captions}
\bigskip
\noindent Fig. 1 - Observational and model results for model B10A for $\beta$ Pic.
Dotted line - model spatial profile at infinite spatial resolution; dashed line -
our KAO point source profile (PSP); solid line - model convolved with PSP; open triangles -
observed average source brightness one-half and one beamwidth off center relative
to the peak observed flux at 47 $\mu$m.  The error bars indicate the combined statistical and calibration
uncertainties.
\bigskip
\noindent Fig. 2 - Model energy distributions for model B10A for $\beta$ Pic relative
to various observations.  Our KAO observations are shown as open triangles
close to the short-dashed line labelled ``FIR'' which is the model
prediction for the flux in our finite size beams as well as for the 800 $\mu$m
data of Zuckerman and Becklin (open triangle also).  The 12, 25, 60 and 100 $\mu$m IRAS
data are shown as open diamonds close to the solid line labelled ``TOT'' which is 
the model prediction for the total flux from the circumstellar disk.  Other points
close to the appropriate model lines for 4'' and 8'' apertures are taken from
BGW.  The error bars indicate the combined statistical and calibration
uncertainties.

\bigskip
\noindent Fig. 3 - Model results for the best fit model described in the text
and Table 4 for Fomalhaut together with various observations.  The spatial scan
panels show our observed 47 and 95 $\mu$m data relative to  the model results,
assuming that our (and the IRAS) data were taken exactly parallel and perpendicular
to the disk axis.  The KAO point-source-profiles are shown as dashed lines.
The energy distribution panel shows the model results, both
for the total flux, and for  that contained within the KAO beam at 47 and 95 $\mu$m
and the 800 $\mu$m beam of Zuckerman and Becklin (dotted).  The IRAS data are also shown
close to (except at 12 $\mu$m) the solid, total flux line.  The error bars indicate the combined statistical and calibration
uncertainties.
\bigskip
\noindent Fig. 4 - Our observational spatial data on the four observed stars in
the Walker and Wolstencroft list (points) relative to the point source profile
(dash-dot line).  None of these stars shows evidence for spatial resolution in
these data.  The error bars indicate the combined statistical and calibration
uncertainties.

\vfill
\eject

\centerline{\bf Authors' Addresses}
\bigskip
\bigskip
\noindent Paul M. Harvey, James Di Francesco, Astronomy Dept., University of Texas, Austin, TX 78712
\bigskip
\noindent Beverly J. Smith, IPAC/Caltech, MS 100-22, Pasadena, CA 91125
\bigskip
\noindent Cecilia Colom\'e, Instituto de Astronomia, U.N.A.M., Apdo. Postal 70-264, DF-04510, Mexico City, Mexico
\bigskip
\noindent Frank J. Low, Steward Observatory, University of Arizona, Tucson, AZ 85721
\vfill
\eject
\end
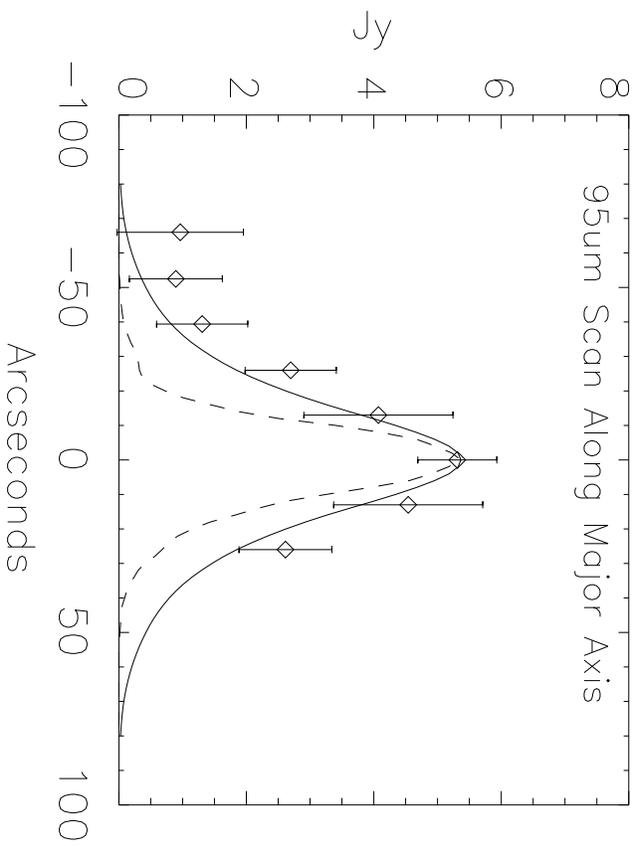
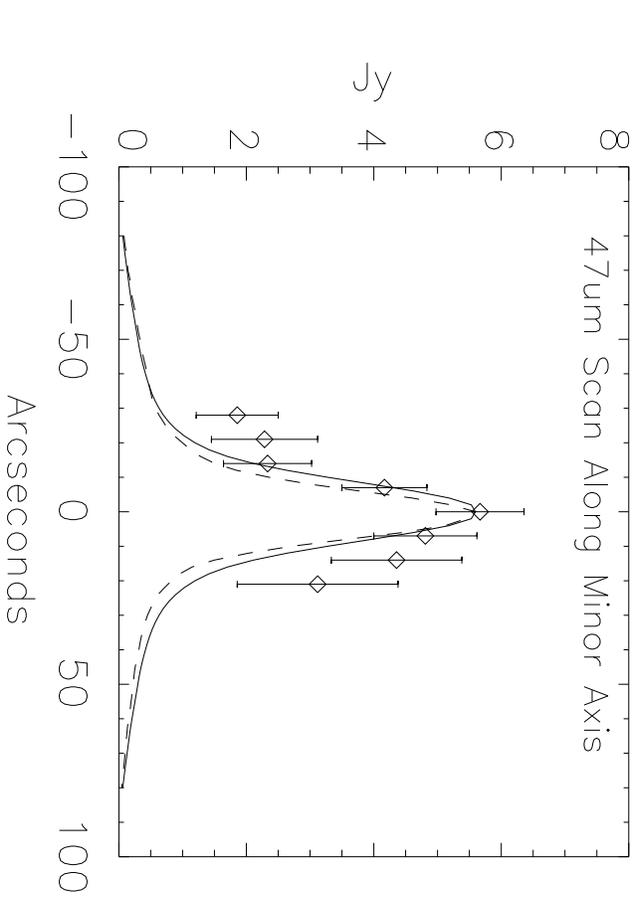
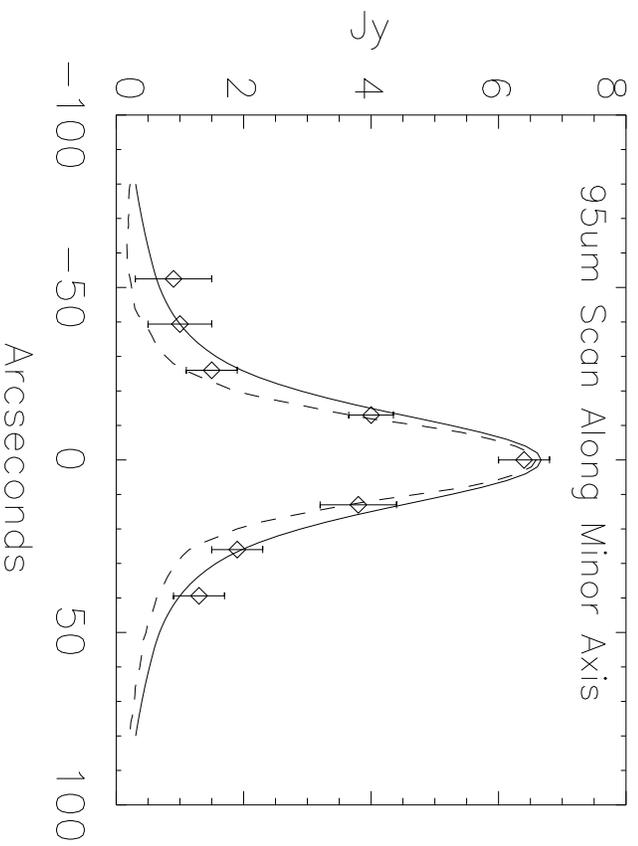
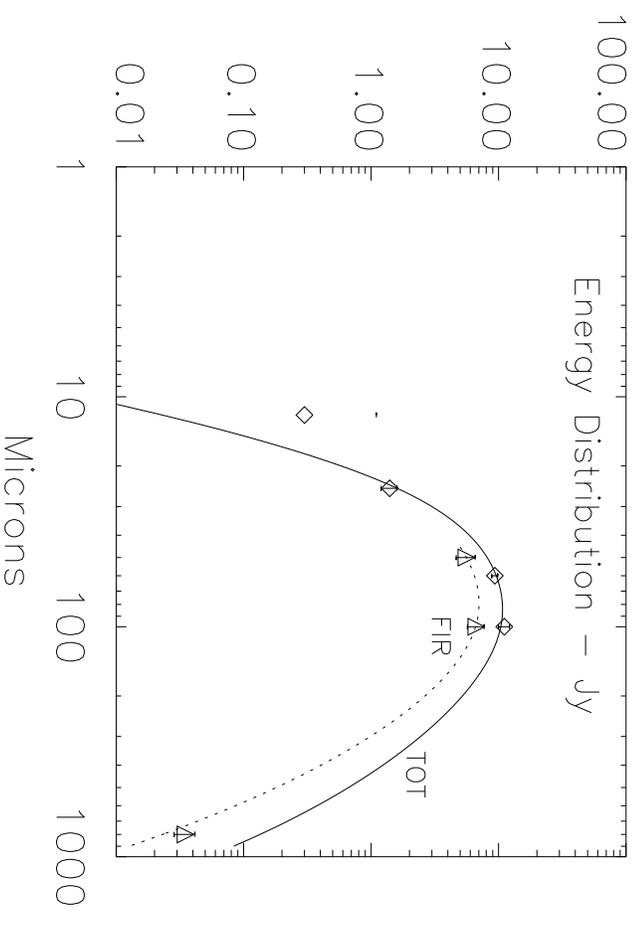
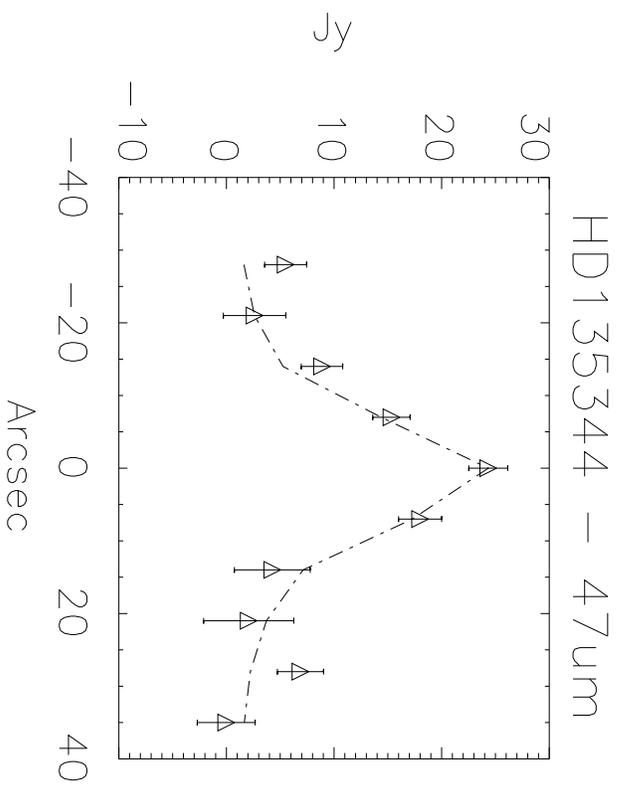
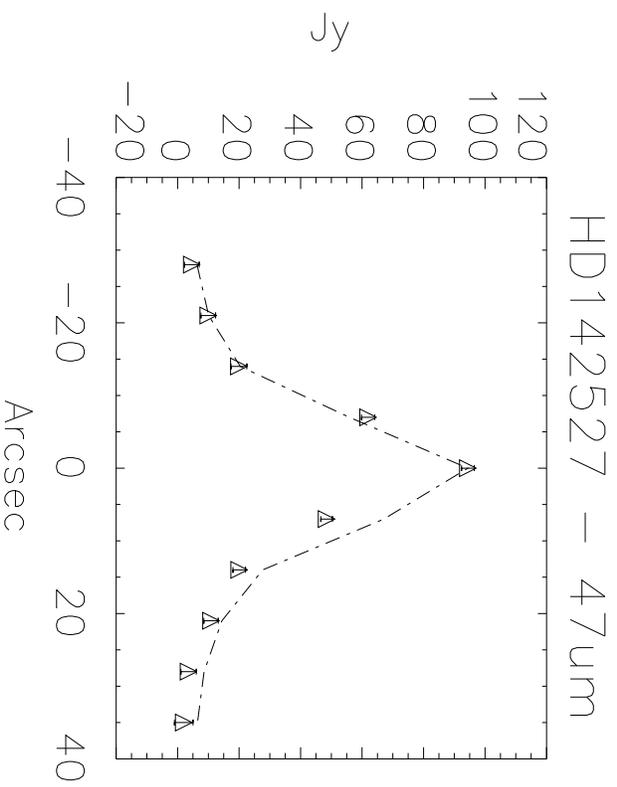
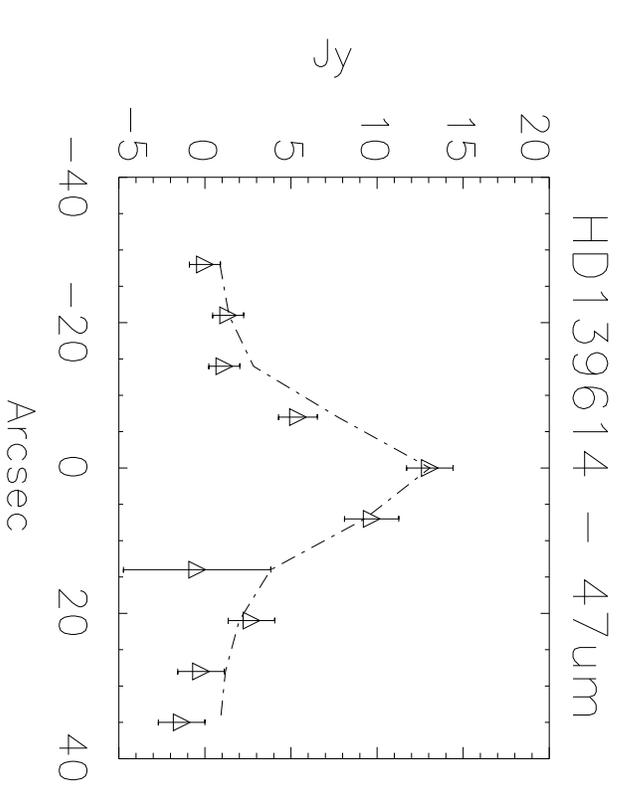
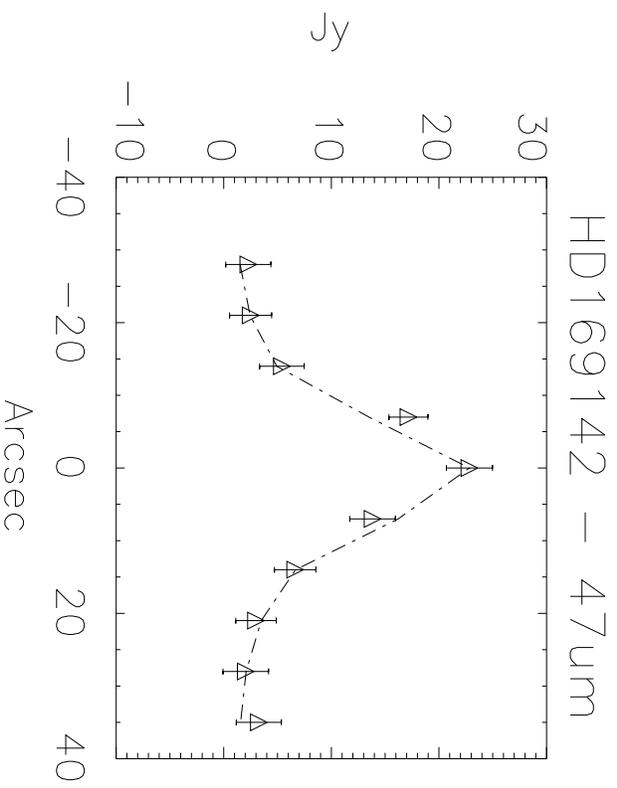
\bye